\documentclass[10pt,twocolumn]{revtex4}
\usepackage{amssymb}
\usepackage{graphicx}
\usepackage{bm}
\pdfoutput=1
\usepackage{epstopdf}
\begin{document}

\title{Glassy dynamics and Landau-Zener phenomena in trapped quasi-one
dimensional coupled Bose-Einstein condensates}
\author{Santiago F. Caballero-Ben\'{\i}tez\footnote{Corresponding author:
scaballero@fisica.unam.mx} and Rosario Paredes 
 }

\affiliation{ 
 Instituto de F\'{\i}sica, Universidad
Nacional Aut\'onoma de M\'exico, Apdo. Postal 20-364, M\'exico D.
F. 01000, M\'exico. }


\date{\today}
\begin{abstract}
The purpose of this article is to address the dynamics of  an
interacting Bose-Einstein condensate confined in coupled one-dimensional
Landau-Zener arrays under the influence of disorder and harmonic confinement. In particular, we
concentrate in studying  the interplay of  disorder and interparticle
interaction on the transfer of atoms depending on the speed of Landau-Zener sweeps.  A dynamical phase diagram summarizing the final situation
across ground state and inverse sweeps is given in terms of the effect of disorder, interaction and the speed of the sweeps.

\end{abstract}

\maketitle

The most important agents that determine the dynamical behavior, and thus the properties, of a quantum system are their structure and the interactions among their constituents. Regarding the influence of the confinement potential, the seminal work by Anderson  \cite{Anderson} investigating the conductance of an electronic system at $T=0$ showed the striking effect of a disordered structure on the change in the conduction in a lattice. On the other hand, the understanding of bosons in disordered media provided by the work of Fisher \cite{Fisher} through the Bose Hubbard model demonstrated the existence of superfluid and Mott insulating phases as a function of the ratio between interactions and hopping strengths, restricting in the Mott insulating case the transport across the media. The structure itself works as an energetic landscape where the atoms or particles in general, are allowed to move under the influence of the interactions. When particles are trapped during their evolution into long-lived inhomogeneous stable or metastable states, a Mott insulating or glass phase can emerge respectively \cite{Carleo}. Apart from the conditions that define the accessible quantum states consistent with the closed system, there are additional factors that may prevent the particles to explore the available quantum Hilbert space in their evolution towards the stationary state. We refer to dynamical restrictions imposed externally during their time evolution. Such protocols are the so called quantum quenches, and can effectively change either the interactions among the system constituents \cite{Sciolla} or the confinement potential. Landau-Zener (LZ) phenomena \cite{Zener} are ideal scenarios where dynamical constrains can be recreated since adiabatically or at a given rate, the energy landscape or {\it potential energy surface} is continuously changed, enhancing the particles to modify its natural behavior.

Even though the intrinsic inhomogeneity created by the harmonic confinement in the experimental realization of ultracold atomic gases, most of the condensed matter and solid state phenomena can be reproduced in the laboratory by adapting the external fields where the atoms move and by adjusting the interactions among the particles in diluted conditions. Particularly, when the atoms are confined in the so called optical lattices, the insulating and glass phases mentioned above can be materialized by modifying the local interactions and setting up a spacial disorder respectively.  Bosons \cite{Damski,Dries,Roth, Roth2,Lewenstein}, fermions \cite{Kondov} and mixtures of them \cite{Ospelkaus,Niederberger} confined in random potentials have demonstrated to exhibit such an insulating and glass phases, being dimensionality another factor participating in the appearance of localization phenomena \cite{Stutzer,Bloch}.
\begin{figure}[t]
\begin{center}
\includegraphics[width=0.47\textwidth]{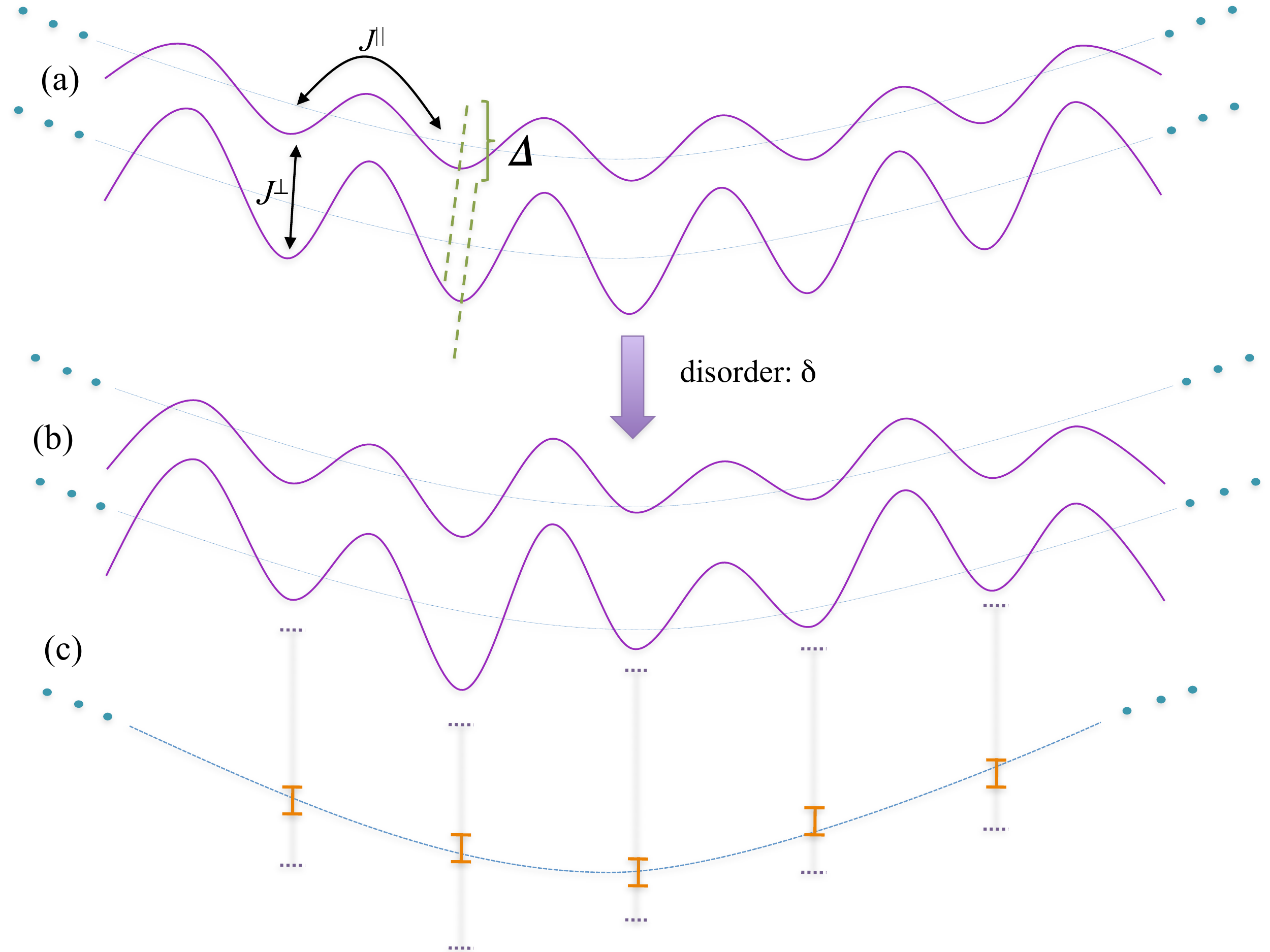}
\end{center}
\caption{(Color online) Schematic representation of a quantum Bose liquid
in chain of double-well traps along a sweep from $0$ to $\Delta$(figures (a) and (b)). Disorder is introduced by changing randomly the potential depth at each site a quantity that varies between what we called strong disorder and weak disorder (dotted and solid lines in figure (c)). 
}
\label{F0}
\end{figure}

The distinctive characteristics that establish the nature of which a quantum phase is present in a Bose fluid at zero temperature are measured in terms of their superfluid fraction, the existence of a gap in the excitation spectra, the behavior of the compressibility, the density and the fixed (integer or semi-integer) number of bosons at each lattice site. In this work we study the many body LZ generalization proposed by Y.-A. Chen \cite{Chen} {\it et al.}  consisting of a Bose Einstein condensate confined in two-coupled chains in 1D where the potential depths defining the lattice sites are linearly modified in time to sweep from an initial energy difference $-\Delta$ to a final difference $\Delta$. In addition to such a time dependent potential, we incorporate the presence of disorder by changing the potential depth at each lattice site.

\begin{figure}[t]
\begin{center}
\includegraphics[width=0.47\textwidth]{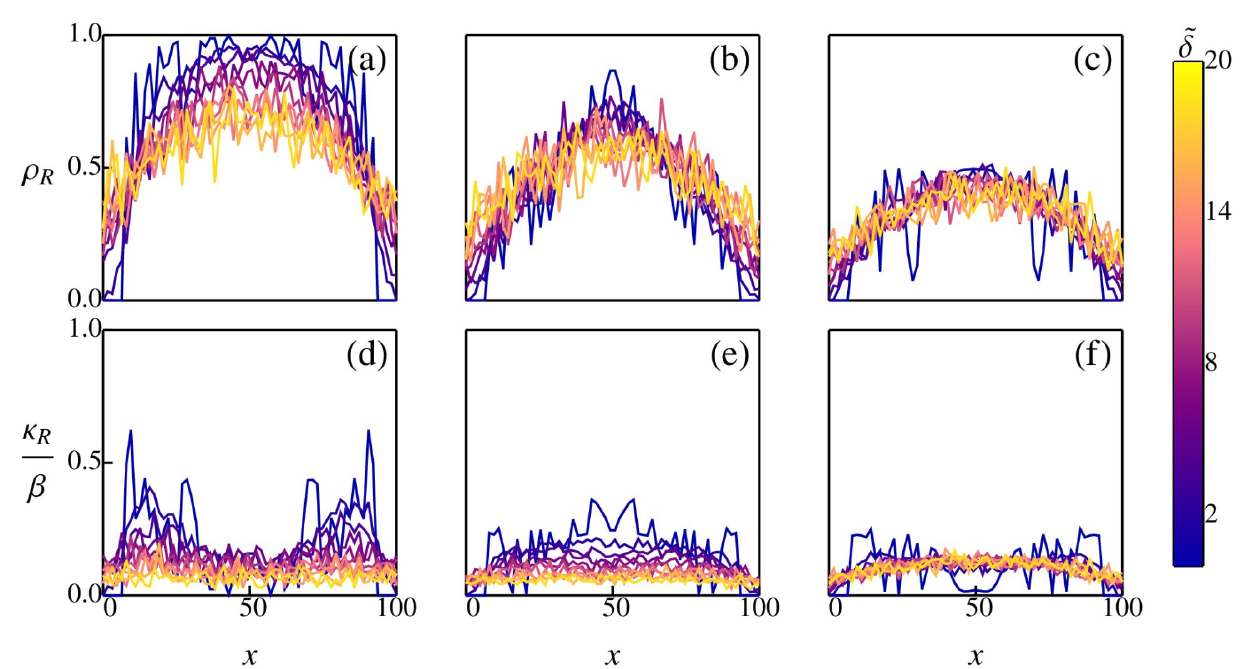}
\end{center}
\caption{(Color online)  The density of transferred atoms to the right
side of the chain ${\rho_R}$ (a-c) and the compressibility
${\kappa_R}$ after a ground state sweep as a function of the
position along the system $x$, for different disorder strengths shown in
the bar on the right, in dimensionless units. Parameters are:
$2\pi/\alpha=10.0$ (adiabatic transfer), $\tilde{V}=15.0/x_0^2$, $x_0=50$, $\tilde{U}=0.1
{\textrm{(a,d)}}, 10.0 {\textrm{(b,e)}}, 20.0  {\textrm{(c,f)}}$.
}
\label{F2}
\end{figure}

\begin{figure}[t]
\begin{center}
\includegraphics[width=0.47\textwidth]{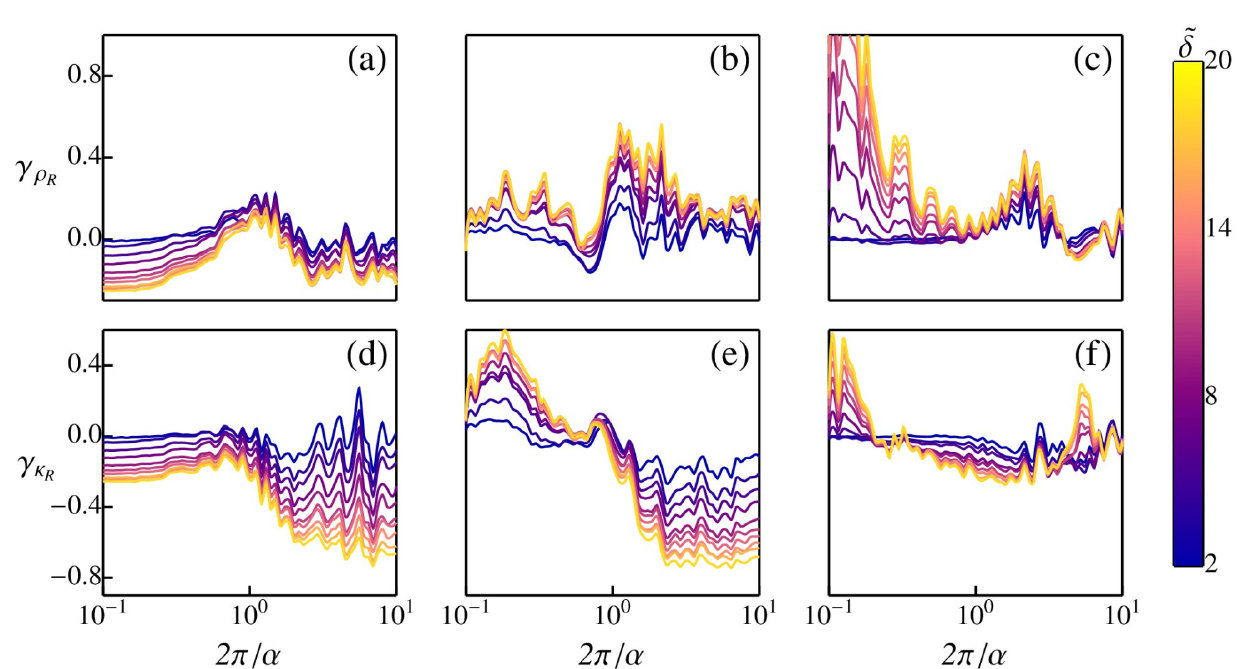}
\end{center}
\caption{(Color online) The average effect along the trap of disorder in
the density of transferred atoms to the right side of the chain
$\gamma_{\rho_R}$ (a-c) and the effect on the compressibility
$\gamma_{\kappa_R}$ after a ground state sweep as a function of the speed
of the sweep $\alpha$, for different disorder strengths shown in the bar on
the right, in dimensionless units. Parameters are:$\tilde{V}=15.0/x_0^2$, $x_0=50$,
$\tilde{U}=0.1 {\textrm{(a,d)}}, 10.0 {\textrm{(b,e)}}, 20.0 
{\textrm{(c,f)}}$.
}
\label{F3}
\end{figure}

First we establish the model that describes the system under consideration. The many body LZ generalization is represented through the Hamiltonian  \cite{Chen, Kasztelan, Caballero}:
\begin{eqnarray}
&&\mathcal{H}^{LZ}=\mathcal{H}^{\parallel}+\sum_{i}\mathcal{H}^{\perp}_i+\sum_{i}\mathcal{H}^{U}_i
\nonumber
\\
&&\mathcal{H}^{\parallel}=-J^{\parallel}\sum_{\nu=R,L}\sum_{<i
,j>}\left(\hat{b}^{\dagger}_{i,\nu}\hat{b}^{\phantom{\dagger}}_{j,\nu}+\textrm{h.c.}\right)
\nonumber
\\
&&\mathcal{H}^{\perp}_i=-J^{\perp}\left(\hat{b}^{\dagger}_{i,R}\hat{b}^{\phantom{\dagger}}_{i,L}+\textrm{h.c.}\right)+\Delta\left(\hat{n}_{i,R}-\hat{n}_{i,L}\right)
\nonumber
\\
&&\mathcal{H}^{U}_i=\frac{U}{2}\sum_{\nu=R,L}\hat{b}^{\dagger}_{i,\nu}\hat{b}^{\dagger}_{i,\nu}\hat{b}^{\phantom{\dagger}}_{i,\nu}\hat{b}^{\phantom{\dagger}}_{i,\nu},
\end{eqnarray}
where $\hat b_{i,\nu}^{\dagger}$ and $\hat b^{\phantom{\dagger}}_{i,\nu}$ are the usual creation and destruction Bose operators with its associated number operator $\hat n_{i,\nu}$. $J^\parallel$ and $J^\perp$ are the intra- and inter-chain coupling energies, and $U$ is the onsite inter-particle interaction strength. While labels $i$ and $j$ designate sites, for $\nu$: $R$ and $L$ are chosen to denote right and left chains respectively. The presence of disorder is introduced by changing in a random manner the potential depth at each lattice site. Such a variation is effectively incorporated in the model by scaling the number of particles with a random number (white noise)  $\delta_{i,\nu}$ in  an interval  $[-\delta,\delta]$ from with mean $0$. Therefore, the Hamiltonian describing disorder at each site has the form,
\begin{equation}
\label{DH}
\mathcal{H}^D_i=\sum_{\nu=R,L}\delta_{i,\nu}\hat{n}_{i,\nu}.
\end{equation}
Similar disorder can be realized experimentally by an speckle potential as in \cite{Aspect,Modugno}.  In our analysis we have considered that disorder is the same in both tubes $\delta_i=\delta_{i,R}=\delta_{i,L}$. The effect of the harmonic confinement associated to the attainment of the condensate is outlined by means of a Hamiltonian term:
\begin{equation}
\mathcal{H}^T_i=\sum_{\nu=R,L}V(x_i-x_0)^2\hat{n}_{i,\nu},
\end{equation}
where $V$ modulates the curvature of the harmonic confinement and $x_0$ denotes the center of the trap, $x_i=(2 x_0/N_x)i$. Thus, the full quantum model considering the effect of the trap and disorder in the LZ scenario is:
\begin{equation}
\mathcal{H}=\mathcal{H}^{LZ}+\sum_i\left(\mathcal{H}^{D}_i+\mathcal{H}^{T}_i\right).
\label{fH}
\end{equation}
\begin{figure}[h]
\begin{center}
\includegraphics[width=0.47\textwidth]{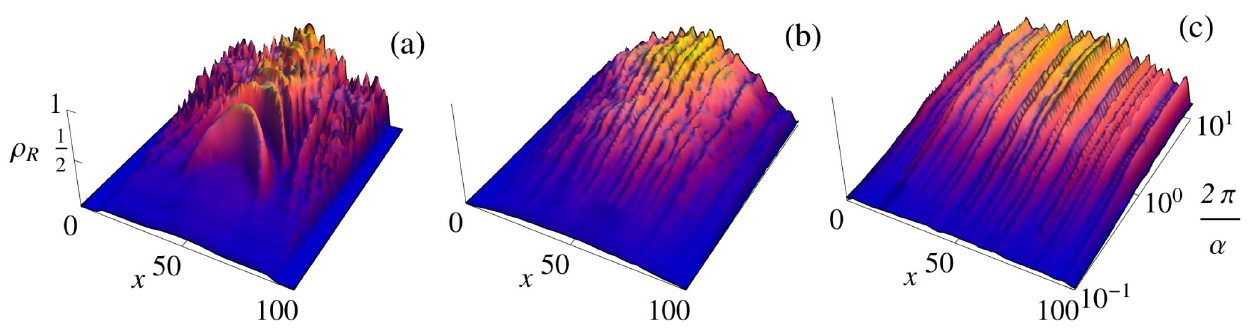}
\end{center}
\caption{(Color online) The density of transferred atoms to the right side
of the chain $\rho_R$  after a ground state sweep as a function of the
speed of the sweep $\alpha$ and the position along the system $x$, in
dimensionless units. Parameters are: $\tilde{U}=10.0$, $\tilde{V}=15.0/x_0^2$, $x_0=50$;
$\tilde{\delta}=0 {\textrm{(a, no disorder)}}, 4.0 {\textrm{(b, weak
disorder)}}, 20.0  {\textrm{(b, strong disorder)}}$, see main text for details.
}
\label{F1}
\end{figure}
Numerical calculations for realistic number of particles become impracticable since Hilbert space scales as $g^{2 N_x}$ and thus dynamical evolution for Hamiltonian (\ref{fH}) requires to diagonalize a matrix of size $g^{4 N_x}$, where  $N_x$ is the number of sites and $g$ is the filling factor. As shown in \cite{Chen}, the number of sites is $\sim$100. Then, to simulate the system with both disorder and the effect of the harmonic confinement, we use the decoupling approximation per site \cite{Caballero}, such that the Hamiltonian becomes a sum of local contributions per site $\mathcal{H}^{LZ}_i$.
\begin{equation}
\mathcal{H}\approx\sum_i\left(\mathcal{H}^{LZ}_i+\mathcal{H}^{D}_i+\mathcal{H}^{T}_i\right)
\end{equation}
where the LZ Hamiltonian is given by,
\begin{equation}
\mathcal{H}^{LZ}_i=J^{\parallel}
\sum_{\nu=R,L}\left(\psi_{i,\nu}\left(\hat{b}^{\phantom{\dagger}}_{i,\nu}+\hat{b}^{\dagger}_{i,\nu}\right)-\psi_{i,\nu}^2\right)+\mathcal{H}^{\perp}_i+\mathcal{H}^{U}_i,
\end{equation}
being $\psi_{R/L}$ order parameters determined self-consistently, such that they minimize the ground state energy of the system for a given set of all the other parameters. The macroscopic state of the system can be written as:
\begin{equation}
|\Psi\rangle\approx|\phi_1\rangle\otimes\cdots\otimes|\phi_{N_x}\rangle
=\otimes_{i=1}^{N_x}|\phi_i\rangle,
\end{equation}
where $|\phi_i\rangle$ is a local many-particle state. The time evolution of the many-particle macroscopic state is given approximately by:
\begin{equation}
|\Psi(t_{n+1})\rangle\approx e^{-i \mathcal{H}(t_n)\delta t}
|\Psi(t_{n})\rangle=\otimes_{i=1}^{N_x}e^{-i \mathcal{H}_i(t_n)\delta t} |\phi_i(t_n)\rangle
\end{equation}
The time evolution is done by first finding self consistently the order parameters that minimize the ground state for all other parameters fixed, where self-consistently the Hamiltonian depends on time linearly via $\Delta(t)=\alpha t$. We do this for each value of $\alpha$, and we proceed with the same calculation as in  \cite{Caballero} , for the time evolution at each site $i$. We should mention that in our calculations all the parameters are referred to the tunnelling coupling parameter $J= J^{\parallel}=J^{\perp}$. Here and henceforth we use the notation $\tilde U = U/J$, $\tilde V= V/J$, $\tilde{\Delta}=\Delta/J$ and $\tilde \delta = \delta/J$. One of the observables that we analyze is the local density of particles in the right tube after a LZ sweep, $
\rho_R(i,\eta,\tilde{\delta})=\langle\tilde\phi_i(\eta,\tilde{\delta})|\hat{n}_{i,R}|\tilde\phi_i(\eta,\tilde{\delta})\rangle/{g},$  for a given set $\eta=\{\alpha,\tilde{U}, \tilde{V}\}$ and disorder amplitude $\tilde{\delta}$ where $|\tilde\phi_i(\eta,\tilde{\delta})\rangle$ is the final state after the LZ sweep. We have considered a filling factor of $g=2$, similar to the experimental conditions of  \cite{Chen}, which means up to two particles per link in the tubes at a given position $x_i$.
In order to have meaningful quantities in terms of the disorder amplitude $\tilde \delta$, we consider sets of $40$ realizations of random numbers for a given disorder amplitude, generating the disorder contribution to the Hamiltonian (\ref{DH}).  After the time evolution is done, we average the realizations for each disorder amplitude generated of the observables $\rho$ and $\kappa$ for each site $i$. A collection of some of these results are shown in Fig. \ref{F1}-\ref{F3}. To support our findings, we go beyond, performing calculations with sets of random numbers with an order of magnitude larger of realizations ($\sim200$)  and we found qualitative agreement with smaller sample sizes. We consider disorder amplitudes in the interval $2\textrm{ (weak disorder)}\leq\tilde \delta\leq20\textrm{ (strong disorder)}$. The LZ sweeps have been made from $t_i=-\tilde{\Delta}_0/\alpha$ to $t_f=\tilde{\Delta}_0/\alpha$ for the ground state sweeps and vice-versa for the inverse sweeps with $\tilde{\Delta}_0=20$. In our simulations we considered $N\sim100$ atoms in the system, with $N_x=64$.

As shown in Fig.\ref{F1} the final density after the sweep inherits the effect of the trapping potential, similar to an inverted parabola depending on the strength $V$. As one increases $V$, less atoms can be contained in the trap.  We also consider the final state to determine the quantum fluctuations in the particle number per site and the local compressibility, 
$
\kappa_R(i,\eta,\tilde{\delta})=\beta(\langle\tilde\phi_i(\eta,\tilde{\delta})|\hat{n}_{i,R}^2|\tilde\phi_i(\eta,\tilde{\delta})\rangle-\langle\tilde\phi_i(\eta,\tilde{\delta})|\hat{n}_{i,R}|\tilde\phi_i(\eta,\tilde{\delta})\rangle^2),
$ with $\beta=k_B T$. 
In overall, we found that as we increase the disorder amplitude $\tilde \delta$ along the trap, the system becomes more homogenous. This comes as no surprise since disorder acts in the diagonal terms in the Hamiltonian, therefore as we increase the disorder amplitude, the atoms see on average less curvature of the harmonic confinement. Thus the system size increases, and the density flattens, see Fig.\ref{F1} and Fig.\ref{F2} panels (a-c). In Fig.\ref{F2} we show the effect of the trapping potential for the state which is approaching the limit of adiabatic transfer  ($\alpha\to0$) for ground state sweeps ($2\pi/\alpha=10$) as a function of the disorder
amplitude for (a,d) weak interaction [$\tilde{U}=0.1$], (b,e) medium interaction [$\tilde{U}=10.0$], (c,f) strong interaction [$\tilde{U}=20.0$]. Consistent with has been analyzed in  \cite{Chen, Kasztelan,Caballero} we found the breakdown of the regular LZ scenario \cite{Zener} where the maximum transfer occurs in the fast sweep limit.  We find consistent behavior with the LZ scenario for inverse sweeps. For weak interaction, disorder destroys the Mott-Insulating (MI) phase occurring in the center of the trap for $\tilde\delta=0$, see panels (a) for the density\cite{Batrouni1} and (d) for the compressibility\cite{Batrouni2}, here quantum fluctuations start to increase
with disorder moving the region of the center of the system away from approximate integer filling. However, as we increase disorder the maximum transfer decreases consistent with localization driven by the Anderson mechanism
\cite{Anderson, Aspect, Modugno}. In panel (d) we can see how the compressibility compensates the effect in the density, due to the curvature away from the center of the trap for weak disorder, this compensation mechanism vanishes as the MI region disappears. As we increase interaction in the adiabatic limit from its medium value (b,e) to the strongly interacting limit (c,f) we find that the effect of disorder starts to saturate and the effect of the interaction prevents the disorder from affecting the system and as one would expect, the compressibility over all the system decreases. Provided that the interaction is below some threshold the system is affected by disorder and the maximum transfer of particles is controlled by the Anderson mechanism in the adiabatic limit, suppressing the transfer locally. However, as we move away from the adiabatic limit, the behavior of the system becomes more interesting. 

Another relevant set of global measures are the spatial averages, $ \bar{\xi}(\eta,\tilde{\delta})=\sum_i\xi(i,\eta,\tilde{\delta})/{N_x},$ where   $\xi$ can be either the
compressibility $\kappa$ or the density $\rho$. Using the above definitions, one can construct the following quantity which measures the
net effect of disorder on average along the system, $\gamma_\xi={\bar{\xi}(\eta,\tilde{\delta}\neq0)}/{\bar{\xi}(\eta,\tilde{\delta}=0)}-1$.

Due to the global
effect of disorder along the trap, a good way of accounting for the effect
of disorder is via $\gamma_{\rho_R}$ and  $\gamma_{\kappa_R}$, which
depending on the speed of the sweep $\alpha$, present non-monotonic
character and are strongly influenced by the many-body interaction, see
Fig. \ref{F3}.  In the weakly interacting regime, Fig.\ref{F3} (a,c),  we
find that in the diabatic limit ($2\pi/\alpha\to 0$), fast sweeps, the
disorder initially suppress both the population transfer and the
fluctuations up to $30\%$. As we slow down the sweeps a maximum occurs (an
enhancement of $20\%$) in the population of right tube, while the effect in
the compressibility vanishes. Slowing down even further returns the system
to have on average a small suppression effect on the density while the mean
compressibility is strongly suppressed. Increasing the interaction, Fig.
\ref{F3} (b,e),  leads to a point where the peak in the effect of the
density sharpens and enhances the average density $\bar\rho_R$ up to $40\%$
for $2\pi/\alpha\sim1$ [here the maximum in $\rho_R\sim0.5$, see Fig.\ref{F2} (b) ], while the effect in the fluctuations with respect
to the absence of disorder is marginal. Afterwards, as the effect in the density
becomes marginal the fluctuations are suppressed, as in the case of weak
interaction. On the other hand fluctuations are strongly suppressed in the adiabatic limit and strongly enhanced for fast sweeps Fig. \ref{F3} (e). However for strong interaction, Fig. \ref{F3} (c,f), the scenario completely changes and the effect of disorder for fast sweeps is to initially enhance the transfer (One must note that although the enhancement  is large the population transfer is exponentially small).  Decreasing the speed suppress the density and then as we
approach the adiabatic limit  the peak in the effect on the density  moves to slower speeds until it disappears. The fluctuations, Fig.\ref{F3} (f) in general are
being suppressed, in agreement with Fig. \ref{F2}. The enhancement or
suppression effect in the density with respect to the speed $\alpha$ has to do with the fact that globally disorder makes more homogenous the system, see Fig. \ref{F1} which
corresponds to $\tilde{\delta}=0.0,4.0,20.0$ and $\tilde{U}=10.0$ and (b,c)
Fig. \ref{F3}.

\begin{figure}[t]
\begin{center}
\includegraphics[width=0.47\textwidth]{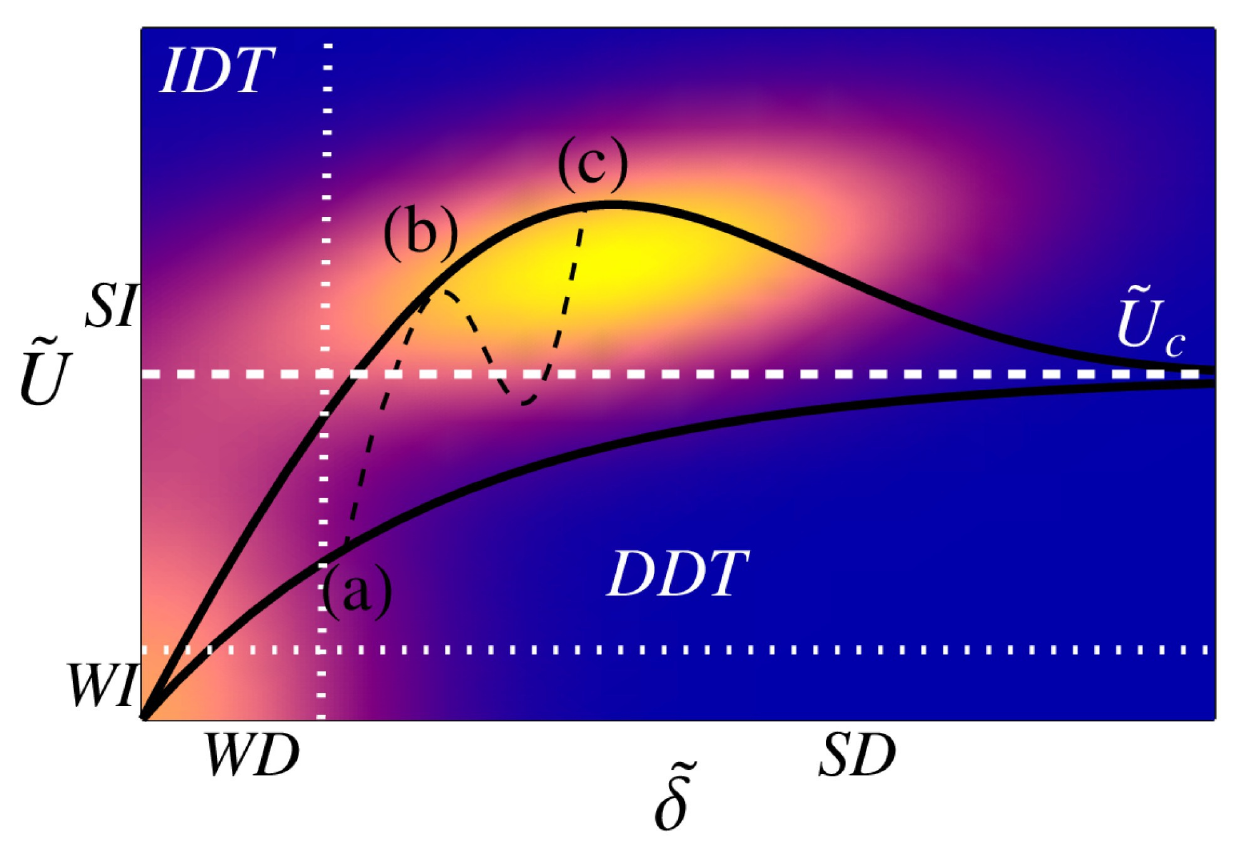}
\end{center}
\caption{(Color online) Qualitative dynamical phase diagram of the system for either ground state or inverse sweeps depending on the interaction strength $\tilde{U}$ and the disorder amplitude $\tilde{\delta}$, units are dimensionless. The boundary between disorder dominated transfer (DDT) and interaction dominated transfer (IDT) is given by the black lines. The behavior concerning the speed of the sweeps is given by the black dotted line, the boundary changes from (a) $2\pi/\alpha\gg1$ [adiabatic limit] to (b) $2\pi/\alpha\sim1$ to (c) $2\pi/\alpha\ll1$ [fast sweeps]. The gray (pink) areas denote the region where the effect on the density is marginal, the brighter area (yellow) denotes the region where enhancement occurs via disorder and the darker areas denote suppression either via disorder (lower right) or via interaction (upper left). The white dots denote the approximate boundary between weakly interacting (WI), strongly interacting (SI), weak disorder (WD) and strong disorder (SD). The white dashed line denotes the critical value of $\tilde{U}_c$ for which interaction dominates and saturates the maximum possible transfer independent of disorder, see Fig. \ref{F2} (c).       
}
\label{F4}
\end{figure}

In the case of inverse sweeps the behavior considering disorder is the opposite with respect to $2\pi/\alpha$ in the ground state sweeps, meaning that for fast sweeps the suppression effect of disorder in the density is stronger for weak interaction, and there exists also a saturation behavior for strong interaction in the maximum transfer (One must note that for inverse sweeps the population transfer in the adiabatic limit is exponentially small). In essence, the adiabatic limit behavior for ground state sweeps and the fast sweeps limit behavior for inverse sweeps are symmetric with respect to the effect of disorder.  We summarize all of our findings in a qualitative dynamical phase diagram in terms of the effect on the mean transfer density, see Fig. \ref{F4}. The net effect of disorder and interaction in the mean density, can be explained in terms of suppression via interaction effects [upper dark region (blue)  in Fig. \ref{F4}] where a saturation effect with respect to disorder occurs. While disorder depending on its strength $\tilde{\delta}$ and the speed of the sweep can have an average suppression effect [lower dark (blue) region in Fig. \ref{F4}] or average enhancement effect [middle bright (yellow) region, in Fig. \ref{F4}] . The local suppression effect [see Fig. \ref{F2}] via disorder is consistent with the Anderson localization mechanism and strongly affects the dynamics, as well as, the effective confinement seen by the condensates of the tubes. Interference effects homogenize the system in the dynamical evolution leading to the picture of  ``glassy dynamics".

This work was partially supported by grant IN108812-2 DGAPA (UNAM).

\end{document}